\documentclass[twocolumn,twoside,slac]{revtex4}
\usepackage{graphicx}
\usepackage{fancyhdr}
\begin{document}

\title{Object Database for Constants: The common CLEO Online and Offline solution}


\author{N. Adam, S. Lee, H. Schwarthoff, T. Wilksen}
\affiliation{Cornell University, Ithaca, NY 14853, USA}

\begin{abstract}
After the successful conclusion of the CLEO III phase, the CLEO experiment at the
Cornell electron positron storage ring CESR is preparing for its transition
to CLEO-c. This new program contains a wide array of Physics studies at $e^+e^-$
collisions at center of mass energies between 3\,GeV and 5\,GeV that will provide
new insights into QCD.
Because the existing Silicon Vertex Detector needed to be replaced within a short time,
a 6 layer Vertex Drift Chamber has been installed in Spring 2003.
At the same time, the existing Ring Imaging Cherenkov Detector, along
with a conventional Drift Chamber, E.M. Calorimeter and Muon Chambers, will
continue to be part of the experiment.

The CLEO Constants Database system must provide efficient access to detector and
analysis constants for a wide variety of purposes, such as hardware configuration,
Online operation and monitoring, calibration, and the complete CLEO analysis framework.
Since the original project, implemented through the Objectivity Object Database
Management System, did not meet those requirements, the system was
redesigned from the ground up, using the same technology. It is currently being
introduced into the production environment without significant interruptions.
In this presentation, we will outline the specifications of the constants database system,
and then report on the process that led to the redesign. Performance comparisons
and insights on relevant design aspects will be shown.

\end{abstract}

\maketitle

\thispagestyle{fancy}


\section{The CLEO experiment}

The 10.6\,GeV $e^+e^-$ collider CESR at Cornell University in Ithaca,
New York, has served as an outstanding Particle Physics research facility
for over 20 years. The measured peak luminosity is
$1.3\times 10^{33} \mbox{cm}^{-2}\mbox{s}^{-1}$.
Its unique detector, constructed and operated by the
CLEO collaboration, has been able to provide a wealth of results from its focus
on B meson studies and a variety of other topics such as charm and tau physics. 
Since the year 2000, the CLEO III detector (\cite{cleoiii}) has
operated with a 4 layer Si vertex detector, and a particle identification
system based on {\em Ring Imaging Cherenkov} (RICH) detectors.

Due to the greatly increased luminosity during the CLEO III phase,
a new readout and computing system was designed to cope with higher
demands (\cite{chep2000},\cite{chep2001}). Throughout the development
process, object oriented principles were applied consequently,
resulting in a coherent and easily extensible set of software
components that now cover all aspects of CLEO computing.
Part of this is the CLEO III database domain, implemented in
an Object Database Management System (\cite{objy}).
Six individual Online databases for alarms, program code,
histograms, run setup data, run statistics data, and constants, comprise
this system. All are implemented as C++ database servers, accessible to
the distributed software components via CORBA communication interfaces
(\cite{corba},\cite{vbroker}).

Two Online databases, for run statistics and constants, are also
used in the CLEO {\em Offline} analysis framework. This framework
is used for all CLEO physics analysis tasks, such as detector
calibration, track reconstruction, Monte Carlo data generation,
and user analysis. It is based on a lightweight on-demand
data delivery approach, as described elsewhere in these proceedings
(\cite{jones}).

\subsection{CLEO-c}

The CLEO III phase has ended in 2003. The storage ring is in the process of
being converted to a charm factory ({\em CESR-c},\cite{cleo-c}),
adapted for center of mass energies between 3\,GeV and 5\,GeV.
It will serve as a new research tool for precision charm quark measurements,
searches for new physics, and $\tau$ lepton studies. It is expected to
provide an array of Quantum Chromo Dynamics (QCD) results that permit
comparison to the most recent generation of lattice QCD calculations
with unprecendented precision.

Since the original facility was not constructed for this energy domain,
a set of superconducting wiggler magnets is installed to ensure
luminosities around $2\times 10^{32} \mbox{cm}^{-2}\mbox{s}^{-1}$.
Several of these devices are already in operation; the upgrade will
be completed in early 2004.
In parallel, the CLEO detector, now called CLEO-c, is undergoing
modifications to allow operation at those lower energies.
Unfortunately, shortly after its deployment, the CLEO III vertex detector
developed severe particle detection inefficiencies. This effect reduced
its usability considerably by causing limitations in secondary
vertex identification. In preparation for CLEO-c, it was replaced
with a 6 layer conventional vertex drift chamber.

\section{The CLEO constants database}

The data structure of a CLEO constants object is determined by the user,
who writes a definition file in c-style syntax, e.g.:
\begin{verbatim}
    Name: RICHChannel
    {
    UInt32 		ChannelAddress
    UInt16 		Crate	SelectBy
    UInt16 		Threshold
    UInt16 		Pedestal
    }
\end{verbatim}
The CLEO build system reads this file and autogenerates a number of
source code files for object database storage, database server
management, CORBA interfacing, client data access, and java classes
for Graphical User Interfaces. This code is compiled and made available
to all higher level software (Online and Offline) in libraries and java archives.

A single constants dataset, called a {\em constants version}, consists
of one of more instances (lines) of data in the form of the described 
structure, such as the four integers in the above example. In CLEO,
some versions have reached ca.~230000 lines.
Every dataset receives a unique {\em version number} when it is
written to the database. In addition, a usage tag and run validity
range can be associated with each constants version.

Users usually do not have write access to the constants database.
Instead, a user can create new constants objects programmatically,
store them as ascii
file, and then submit them to the central constants management system
for installation.  After applying some consistency checks and safeguards,
that system writes the new data to the database and updates validity
information.

After 2 years of operations, the CLEO constants database had proven to be
a reliable design that was extendable for new data types. It became clear
that separate copies of the full database were needed for Online operation,
track reconstruction processing, public analysis, and remote site
Monte Carlo generation. For that purpose, the protected Online database
was designated to be the {\em master database}, i.e. the only existing
copy available for write access. A number of copies acted as {\em slave
mirrors}, readonly versions that were updated from the master as needed.
This scheme allowed minimization of the load on the master, a requirement
for smooth and independent Online detector operation.

\section{The constants database redesign}

Despite the success in some areas, experience revealed that the
constants database system was designed in a way that did not
match the predominant usage patterns. Performance degraded
with increased database size, causing concern for Online operation.
Hard disk and CPU contention on the server nodes could lead to
wait times of minutes for a single request. The system could not
be expected to scale well for the rising number of clients.
This situation led to the decision to rewrite all constants
database access code to ensure future scalability and efficiency.

\subsection{Redesign - why?}

A data object that is stored in an object database requires a
unique identifier (object identifier, OID) for data
management within the database and for retrieval. Such an
object can have almost any shape, typically expressed in a c-style
structure (or a C++-style class). A certain overhead for OID
and storage is necessary and typically amounts to 20--30 Bytes,
in addition to the space needed by the object itself.
Retrieving an object also entails a certain computing overhead
that is mostly independent of object size. From this fact
we conclude that it is desirable to keep the number of objects
retrieved in a single database transaction small.

This notion was not part of the original constants database
design: A CLEO constants version, with a structure as described
in the above example, could have many lines. The original system
architects had implemented this as a series of many OIDs,
one per line. The consequence was considerable overhead in
storage space and CPU time needed for retrieval:
a download request always asked for a complete constants version,
i.e. many OIDs. In addition, an inefficient search algorithm that
often had to iterate through a large number of objects
contributed to a steady decline in download performance
as data volume rose over time.

Another limitation became relevant when new versions for
one of the existing data types were created in greater numbers
than originally foreseen. Since every object must have a unique
OID, the number of objects in any part of a database is limited.
In this case, there could not be more than $2^{15}-1 = 32767$ versions.
A workaround was possible, but only at the
expense of more complicated data distribution and longer search times.

In the data organization domain, the original system violated
a basic rule of database design: Data normalization: the data contents
of a constants version was stored not only together with its version
number, but also with its usage information (usage tag and run validity
range). In such a setup, one is severely limited in how data access
can be managed. It is, for example, not possible to associate more than one
validity range - usage tag combination efficiently with an object, since
a search with a certain tag and run number requires examination of many
data objects. Since CLEO required several such combinations, a cumbersome
set of separate structures had to be invented to provide this functionality.
The maintenance of those structures was particularly arduous.

\begin{figure*}[t]
\centering
\includegraphics[width=135mm]{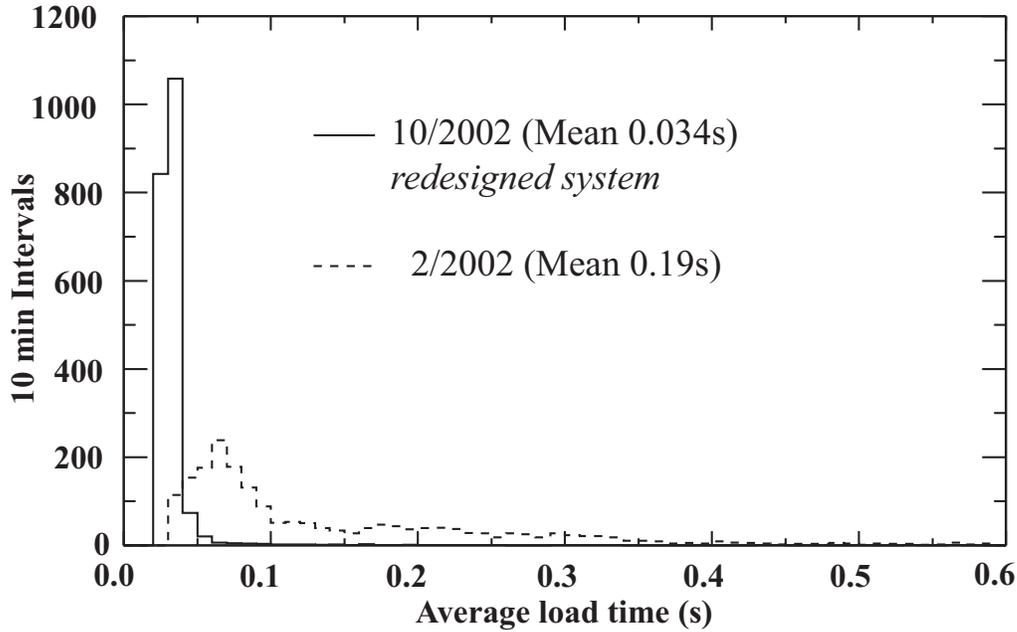}
\caption{Average load time for a single constants version,
measured over 10 min intervals.
The server side CORBA conversion overhead is included.
} \label{fig:averagetimes}
\end{figure*}

The combination of the described shortcomings eventually provided
a strong justification for the redesign of the database structure.

\subsection{The new constants database system}

Motivated by a wealth of experiences from the CLEO Online and
Offline production systems, the specifications for the redesign
could be determined with ease:

\begin{itemize}
\item One OID per constants version
\item No linear search on retrieval
\item Copious OID space for a large number of versions
\item Constants data stored independently of usage information
\item Navigational elements that allow flexible usage
      management incorporated into the basic design
\end{itemize}

According to these requirements, the new design consolidated
all lines of one constants version into a single database object.
This approach reduces a database query to the retrieval of one
object; download pattern and storage structure are exactly matched.
The new organization now permits very fast data retrieval through
a set of management structures that can be traversed quickly.
A lookup with a usage tag and run number is organized so that
a validity list is searched in a tag dictionary, which then
holds a direct reference to the desired data object.
The need for linear object searches was completely eliminated by
using indexed version lists and hash maps for fast lookup via
dynamic usage tags. Loading an object with a specified version number
is as easy as opening a version list and dereferencing it
with the version number as array index.

With the number of database objects greatly reduced, the
available OID space is now considerably larger than needed
for the lifetime of the CLEO experiment.
At the same time, the space requirements are less than half of
the original needs. A remarkable success story of object
oriented client-server design was also the fact that the
client interfaces did not have to be modified at all, since
the redesign could remain completely confined to the
database server implementation. This allowed a smooth
transition of all running build and production systems.

For an illustration of the performance improvement that
was achieved through the redesign, see Figure~\ref{fig:averagetimes}.
While the average load time of a single constants version
($\sim$\,0.2\,s), requested by regular physics reconstruction and
analysis programs, were already an improvement from intermediate
changes, the new system was able to reduce this time by more than
a factor of five. The distribution is much narrower,
which translates into a fast, responsive, and reliable
behaviour. Since a single analysis job can request over
100 different constants versions, the impact on real world
operation is dramatic.

\subsection{Experiences with commercial software}

The generic trend in the High Energy and Nuclear Physics
computing community to purchase commercial software for
their ever growing production systems has generated
many varied experiences. Physicists do not write their own
CORBA implementations, but they have realized that their
database needs may be different from what industry
requires, and that developing solutions on their own
is possible and may turn out to be the better way.
Of concern for the entire community is the fact that the widely used
commercial implementation of an Object Database Management
System by Objectivity Inc. (\cite{objy}) is being deprecated
by almost all major collaborations for their large scale
data storage needs. A variety of reasons is cited:
\begin{itemize}
\item The Physics community is reluctant to adopt software
without access to the source code, since its computing systems
are usually highly non-standard implementations that require
customization.
\item An object database does not match traditional data
access and usage patterns.
\item Large scale data storage with an object database
requires a highly sophisticated and finetuned setup and
time consuming expert operation.
\item Financial means in public research are limited.
\end{itemize}
Although those issues are of little direct consequence for the 
project described here, because of the much smaller database
sizes, a decline in overall acceptance will have an impact
on product pricing, availability, and support.

Similarly, the situation continues to be challenging with regard
to platform support. Third party software packages, such as CORBA
implementations, make inroads into the Linux world at a slow pace.
But the Linux on Intel platform is in the process of displacing
all other hardware/operating system combinations in the field.
This situation has placed limitations on what can be done with
commercial software, and may prove a confining factor in future
upgrades of the CLEO computing systems.

\begin{acknowledgments}
This work has been supported by the US Department of Energy, the National
Science Foundation, and the Alexander von Humboldt foundation.
\end{acknowledgments}


\end{document}